\documentclass[10pt]{article}
\pdfoutput=1
\usepackage[utf8]{inputenc}
\usepackage[T1]{fontenc}
\usepackage[pdftex]{geometry}
\usepackage[english]{babel}
\usepackage[authoryear]{natbib}
\usepackage{textcomp,amsmath,amssymb,amsfonts,pdflscape,longtable,afterpage,calc,booktabs}
\usepackage{url}
\usepackage{graphicx}
\begin{document}
\let\WriteBookmarks\relax
\def\floatpagepagefraction{1}
\def\textpagefraction{.001}
\thispagestyle{empty}
\begin{center}
\Large Thomas Ruedas\textsuperscript{1,2}\\[5ex]
\textbf{Comment on ``The SPOCK equation of state for condensed phases under arbitrary compression'' by R.~Myhill}\\[5ex]
final version\\[5ex]
02 June 2025\\[10ex]
published as:\\
\textit{Geophysical Journal International} 242(2), doi:10.1093/gji/ggaf210 (2025)\\[15ex]
\normalsize
\textsuperscript{1}Museum für Naturkunde Berlin, Leibniz Institute for Evolution and Biodiversity Science, Germany\\
\textsuperscript{2}Institute of Space Research, German Aerospace Center (DLR), Berlin, Germany\\
\rule{0pt}{12pt}
\end{center}
\vfill
\footnotesize The version of record is available at \url{http://dx.doi.org/10.1093/gji/ggaf210}.\\
This author pre-print version is shared under the Creative Commons Attribution License (CC BY 4.0).
\normalsize
\title{Comment on ``The SPOCK equation of state for condensed phases under arbitrary compression'' by R.~Myhill}\\[5ex]
\author{Thomas Ruedas\textsuperscript{1,2}\thanks{Corresponding author: \texttt{Thomas.Ruedas@mfn.berlin}}}
\date{}
\maketitle

\begin{abstract}
It is shown that the SPOCK equation of state is equivalent to the Variable Polytrope Index equation of state.
\end{abstract}

\noindent In a recent paper, \citet{Myhill25b} introduced the ``Scaled Power Of Compression for K-prime'' (SPOCK) equation of state (EoS) for condensed materials under arbitrary compression and demonstrated its validity and excellent performance even to extreme pressures while obeying thermodynamical constraints that many frequently used EoSs fail to satisfy. However, it turns out that the SPOCK EoS is a rediscovery with an independent and arguably more concise derivation of the Variable Polytrope Index EoS presented by \citet{Wepp:etal15}. In the following, the equivalence of some essential relations in both papers is shown.

\citet{Myhill25b} expressed the thermodynamic functions in terms of the specific volume $V$ or indeed a logarithmic measure of compression, $f$, that is similar to the Hencky strain as used by \citet{PoTa98} in their Logarithmic EoS. By contrast, \citet{Wepp:etal15} used the density $\varrho$ as the independent variable. The relation between them is
\begin{equation}
    f\equiv\ln\left(\frac{V}{V_0}\right)=\ln\left(\frac{\varrho_0}{\varrho}\right),\label{eq:f}
\end{equation}
where the subscript 0 indicates the reference state.

Starting from the definition of the bulk modulus, $K$, it and its pressure derivative $K'$ can then be expressed in terms of $f$:
\begin{align}
    K&=\left(-\frac{1}{V}\frac{\partial V}{\partial P}\right)^{-1}=-\frac{\partial P}{\partial\ln\left(\frac{V}{V_0}\right)}=-\frac{\partial P}{\partial f}\label{eq:K}\\
    K'&\equiv\frac{\partial K}{\partial P}=\frac{\partial f}{\partial P}\frac{\partial K}{\partial f}=-\frac{\partial\ln K}{\partial f}\label{eq:Kp}
\end{align}
\cite[Eqs.~2 and 3]{Myhill25b}, with $P$ designating the pressure.

\paragraph*{Pressure derivatives of the bulk modulus.}
Both authors used the first pressure derivative of the bulk modulus, $K'$, as the foundation of their EoSs; \citet{Wepp:etal15} demonstrated that $K'$ ($B'$ in the notation used in their paper) corresponds to the polytrope index $n$ in the polytrope EoS $P=C\varrho^n$, where $C$ is a constant. The polytrope EoS is frequently used in astrophysical contexts to describe the properties of matter in stars.

In their Eq.~34, \citet{Wepp:etal15} proposed the form
\begin{equation}
    K'=\frac{\mathrm{d}K}{\mathrm{d}p}=A_0\left(\frac{\varrho_0}{\varrho}\right)^{A_1}+A_2\label{eq:Kp-W15}
\end{equation}
for the pressure derivative of the bulk modulus, with $A_0$, $A_1$, and $A_2$ being constants. Substituting Eq.~\ref{eq:f} into it directly yields the SPOCK form of $K'$,
\begin{equation}
    K'=ab\exp(af)+c\label{eq:Kp-M25}
\end{equation}
with constants $a$, $b$, and $c$, as given in Eq.~8 of \citet{Myhill25b}. Comparison of the constants in Eqs.~\ref{eq:Kp-W15} and \ref{eq:Kp-M25} leads to the relations
\begin{align}
    a&=A_1=-\frac{K''_0 K_0}{K'_0-K'_\infty}\label{eq:a}\\
    b&=\frac{A_0}{A_1}=-\frac{(K'_0-K'_\infty)^2}{K''_0 K_0}\\
    c&=A_2=K'_\infty.\label{eq:c}
\end{align}
$K'_\infty$ is the pressure derivative of the bulk modulus at infinite compression.

Both authors discussed the need to make assumptions on $K''_0$ and referred to the heuristic relation
\begin{equation}
    K''_0=-\frac{K'_0}{K_0}\label{eq:Kpp0}
\end{equation}
\cite{BoRoBoRo06,HoPo11}. \citet[Eq.~B14]{Wepp:etal15} adopted this relation directly, whereas \citet[Eq.~32]{Myhill25b} advocated the condition $-K''_0 K_0\geq K'_0-K'_\infty$ as a more flexible constraint that includes Eq.~\ref{eq:Kpp0} because of a strict theoretical lower limit for $K'_\infty$ of 5/3 \cite{Stacey00}.

\paragraph*{Bulk modulus.}
Integration of Eq.~\ref{eq:Kp-W15} yields
\begin{equation}
    K=K_0\exp\left\lbrace\frac{A_0}{A_1}\left[1-\left(\frac{\varrho_0}{\varrho}\right)^{A_1}\right]\right\rbrace \left(\frac{\varrho}{\varrho_0}\right)^{A_2}
\end{equation}
\cite[Eq.~B8]{Wepp:etal15}. Substitution of Eqs.~\ref{eq:f} and \ref{eq:a}--\ref{eq:c} leads directly to the SPOCK version of the bulk modulus,
\begin{equation}
    K=K_0 \exp b \exp[-b \exp(af)-cf]
\end{equation}
\cite[Eq.~15]{Myhill25b}, which was derived there by integrating Eq.~\ref{eq:Kp-M25} using Eq.~\ref{eq:Kp}, and exponentiating.

\paragraph*{Pressure.}
Integrating the definition of the bulk modulus written as a function of density from 0 to $P$, \citet[Eq.~35]{Wepp:etal15} found an expression for the pressure as a function of $\varrho$:
\begin{equation}
    P=\frac{K_0 \exp\left(\frac{A_0}{A_1}\right)}{A_1}\left\lbrace\left(\frac{\varrho}{\varrho_0}\right)^{A_2} E_\frac{A_1+A_2}{A_1} \left[\frac{A_0}{A_1}\left(\frac{\varrho_0}{\varrho}\right)^{A_1}\right]-E_\frac{A_1+A_2}{A_1} \left(\frac{A_0}{A_1}\right)\right\rbrace.\label{eq:p-W15}
\end{equation}
The $E$ terms in this equation are the generalized exponential integral
\begin{equation}
    E_m(x)=\int\limits_1^\infty \frac{\exp(-xt)}{t^m}\mathrm{d}t=x^{m-1}\Gamma(1-m,x),\label{eq:GEI}
\end{equation}
where $\Gamma(1-m,x)$ is the incomplete Gamma function
\begin{equation}
    \Gamma(\alpha,x)=\int\limits_x^\infty \frac{t^{\alpha-1}}{\exp t}\mathrm{d}t
\end{equation}
\cite[Eqs.~6.5.3 and 6.5.9]{AbSt64}. Expressing the generalized exponential integrals in terms of the incomplete Gamma function by Eq.~\ref{eq:GEI}, Eq.~\ref{eq:p-W15} becomes
\begin{equation*}
    P=\frac{K_0 \exp\left(\frac{A_0}{A_1}\right)}{A_1}\left(\frac{A_0}{A_1}\right)^\frac{A_2}{A_1}\left\lbrace \Gamma\left[-\frac{A_2}{A_1},\frac{A_0}{A_1}\left(\frac{\varrho_0}{\varrho}\right)^{A_1}\right]-\Gamma\left(-\frac{A_2}{A_1},\frac{A_0}{A_1}\right)\right\rbrace.
\end{equation*}
Using the shorthand
\begin{equation*}
    \Gamma(\alpha,x_1,x_2)\equiv\Gamma(\alpha,x_1)-\Gamma(\alpha,x_2)=\int\limits_{x_1}^{x_2} \frac{t^{\alpha-1}}{\exp t}\mathrm{d}t
\end{equation*}
and substituting once more Eqs.~\ref{eq:f} and \ref{eq:a}--\ref{eq:c} yields
\begin{equation}
    P=K_0\frac{\exp b}{a} b^\frac{c}{a} \Gamma\left[-\frac{c}{a},b\exp(af),b\right],
\end{equation}
which corresponds exactly to Eq.~21 in \citet{Myhill25b} if the reference pressure $P_0$ is 0. \citet{Myhill25b} derived this relation by integration of Eq.~\ref{eq:K}.

\paragraph*{Extensions.}
The SPOCK/Variable Polytrope Index EoS is an isothermal EoS. It is worth noting that \citet{WiHo20a}, who referred to the latter as the ``Varpoly EoS'', developped a Mie--Grüneisen--Debye-like thermal extension and also gave consideration to expanded states ($\varrho <\varrho_0$), which are relevant in particular for the rarefaction wave that follows the shock in a hypervelocity impact.

\subsubsection*{Data Availability}
There are no new data associated with this article.


\end{document}